\documentclass[adp,fleqn]{w-art}

\RequirePackage{latexsym}
\RequirePackage[english]{babel}
\RequirePackage{amssymb}  
\RequirePackage{amsbsy}   
\RequirePackage{amsmath}  
\usepackage[T1]{fontenc}
\usepackage[dvips]{graphicx}

\newcommand{\pp}{$p\mspace{2mu}$-$p$ }
\newcommand{\ud}{\mathrm{d}}
\newcommand{\ee}{$e^+e^-$ }

\newcommand{\lead}{$\mathsf{L}$}   
\newcommand{\effl}{leading effect}
\newcommand{\efflc}{Leading Effect}
\newcommand{\atot}{$A_{\text{TOT}}$ }
\newcommand{\mx}{$M_{\text{X}}$ }
\newcommand{\meff}{$M_{\text{eff}}$ }
\newcommand{\rads}{$\sqrt{s}$ }
\newcommand{\ltot}{$\mathcal{L}_{\text{TOT}}$ }
\newcommand{\qht}{$M_{\text{X}}\equiv \sqrt{(q_{\textsc{tot}}^{\textsc{had}})^2}$}
\newcommand{\eneff}{effective energy}
\newcommand{\cms}{\textsc{cms}}
\newcommand{\nch}{$\big< n_{ch} \big>$ }
\newcommand{\aqsc}{Available Quadri-Scalar}
\newcommand{\aqs}{available quadri-scalar}
\newcommand{\eht}{$E^{\textsc{had}}_{\textsc{tot}}$ }

\newcommand{\qlt}{$M_{\text{L}}\equiv \sqrt{(q_{\textsc{tot}}^{\textsc{lead}})^2}$}
\newcommand{\te}[1]{\textsc{#1}}
\author{Corrado Appignani\footnote{e-mail: {\sf
appignani@bo.infn.it}}}  
\address{Dipartimento di Fisica --
Universit\`a di Bologna -- Via Irnerio 46 -- I-40126 --
Bologna -- Italy}
\title{QCD universality recovered via the Total Available Quadri-Scalar}

\begin{document}
\Receiveddate{23 July 2003}
\Reviseddate{}
\Accepteddate{12 March 2004 by G.~R{\"o}pke}
\Dateposted{}


\begin{abstract}

After presenting a brief review of the phenomenology of the Leading
Effect, we define a new variable, the ``Total Available
Quadri-Scalar'' (\atot\!\!) and propose it as the invariant quantity
effectively available for the production of the multihadronic final
states. The introduction and definition of this new varaible are
justified by means of simple geometrical-kinematical considerations
and we show that \atot reduces to the so-called \eneff\ in the single
specific situation where the use of the latter applies. Using \atot
to re-plot existing data, the quantity \nch is shown to be a
``Universality Feature'' -- that is, independent from the colliding
particles, the collider nominal energy, and even from the hadronic
invariant mass -- as imposed by QCD universality.

\end{abstract}

\maketitle

\section{Introduction}

Basically speaking, QCD shows two main features. One of them is perturbative in
nature and is Asymptotic Freedom. This property is understood in terms of a
negative $\beta$ function, which implies that the QCD gauge coupling
$\alpha_\te{S}$ decrease with an increasing $q^2$. The other, Confinament, is
non-perturbative and has not received, up to now, a satisfactory theoretical
explanation.  

Another important but almost forgotten feature of QCD is the Effective
Energy.
This is also a non-perturbative effect and is, roughly speaking, the mechanism
through which the initial (nominal) energy is shared among different processes,
one of which is the hadronization. The slice of energy 
that goes to the latter is not the nominal, but an effective energy. It is this
one, and not the whole nominal energy, that is at disposal for particle
production, as explained in the paper in a more datailed  way.

This very interesting property is practically no more studied, even
if it could shed some light on and provide an alternative approach to
the question of how the hadronization mechanism works. The
fundamental idea that lies behind the \eneff\ approach is that of
distinguishing two main phases inside any given interaction, namely,
the quantum number conservation (or flow), and the hadronization in
the proper sense of the word, that is the process by which quarks and
gluons ``hadronize'' and become observable matter (hadrons). 

Some papers published in the early '80s \cite{evid,nc2,nc3,nc4}
showed that, whatever the invariant quantity available to the
hadronization process, it should equate the total hadronic energy of
the whole system as evaluated in the CMS (therefrom the name
\eneff).

In fact, if a whole set of quantities (specified in the next section)
are studied in terms of that portion of energy, the plots relative to
different kinds of collisions between different kinds of particles
converge toward the same curve.

Despite the great importance of such a result, neither the correct
invariant representation has been identified, nor a satisfactory
justification for the choice of the quantities adopted in the past to
represent it has been given. 

In fact, the quantity commonly used to plot data and to study the
world-widely collected experimental results is \qht\  but, as will be
showed in this work, this quantity cannot work if we insist in
separating the aforesaid phases of the interaction. Furthermore, as
some recent papers show, (see e.g.~\cite{osaka}) \mx\ is unable to
reveal the Universality Features when also DIS processes are taken
into consideration.

Purposes of this paper are to introduce, by means of simple
kinematical considerations, what is believed to be the correct
Lorentz--invariant representation of the quantity from now on called
the ``\aqsc'' and to show how it is possible, by correcting the above
mentioned result, to let the universality features be ``revealed
again''.

The outline of this paper is as follows: in Section 2 a brief review
of the phenomenology of the \efflc\  is presented; in Section 3 a new
variable, \atot\!, is conceptually introduced while its formal
definition is given in Section 4. In Section 5 the relations between
\atot and \mx are examined. Evidence of \nch vs \atot universality is
provided in Section 6 with the final result given in Section 7. In
Section 8 rescaling of \pp data is discussed and in Section 9 a brief
analysis of the dependence of our result from kinematical condition
is given. Section 10 contains the conclusions. 

\section{Phenomenology of the Leading Effect}

The fact that the total energy available to particles production in a
given type of interaction is not, in general, the nominal energy of
the reaction, but another quantity that takes into account the \effl,
is one of the most important discoveries made in the early '80s
\cite{creation}. Before that time, all the measured quantities were
analysed in terms of \rads and this brought to different results in
different experiments. This was commonly accepted even if it was in a
flagrant contrast with the QCD universality.

In fact, at a fundamental level, the final state of any interaction
should depend only on some (Lorentz-invariant) scalar variable
believed to be available to the hadronization mechanism, but not on
how that quantity has been put together. This means that, had this
quantity been \rads\!, the results of any analysis should have been
the same, independently on the kind of reaction under exam (e.g.
\!\ee\!\!, \pp\!, DIS). Instead, the results were all different and
there was no explanation for this situation, to which people referred
as ``the hidden side of QCD''.

As mentioned, in the early '80s it was pointed out \cite{evid} that
QCD universality could have been made explicit if the quantity
\eht\!, that is, the total hadronic energy of the reaction given by
subtracting the energy of the leading particle(s) from the nominal
energy of the reaction, was used to plot data instead of \rads\!\!.
Within a few years, this effect, called the \efflc, would have been
shown to be universal: in fact, no matter if the interaction studied
was strong, electromagnetic or weak, the \effl\ was always present
\cite{creation,nc7,nc10,hadmes}.

\eht was soon after called the ``Effective Energy'', the name
claiming for \eht to be the portion of energy effectively available
to the production of the multihadronic systems present in the final
state, after subtraction of the energy of the leading particle.

The leading particle was defined to be the particle leaving the
interaction vertex with the highest longitudinal momentum
\cite{evid}. The role of this particle was to carry, totally or
partially, the quantum numbers (as $J^{PC}$ or flavour) from the
initial to the final state. The transfer of these quantum numbers
from the reacting particle(s) to the leading particle was called the
Quantum Number Flow (hereafter: QNF).

From 1980 to 1984 a series of experiments were conducted at the ISR
(CERN) to establish if \eht was the effective energy. All these
experiments proved that there were no  differences between \pp and
\ee collisions results if \eht was used to perform the analyses. The
quantities measured in these experiments were called the
``Universality Features'' as they showed the same behaviour whatever
the kind of experiment and the nominal energy of the collider. 
Some of them are:
\begin{enumerate} 
\renewcommand{\labelenumi}{ \arabic{enumi} ) \ }
\item ${\displaystyle \big< n_{ch}\big>\ =\ }$ Mean charged particle 
multiplicity ~\cite{nc3}\\[-2mm]
\item ${\displaystyle \frac{\ud\sigma}{\ud x_{\textsc{r}}}}$ = 
\ Fractional energy distribution ~\cite{evid,nc2}\\[-2mm]
\item ${\displaystyle \frac{\ud\sigma}{\ud p^2_t}}$ = 
\ Transverse squared momentum distribution ~\cite{nc9}
\item ${\displaystyle \frac{\ud\sigma}{\ud(\frac{p_t}{<p_t>})}}$ = 
\ Reduced transverse momentum distribution ~\cite{nc8}
\item ${\displaystyle \alpha\equiv\frac{E_{ch}}{E_{\textsc{had}}}}$ = 
\ ``Charged'' energy ~\cite{nc5}
\item ${\displaystyle \frac{\ud\sigma}{\ud \big< p_T^2 \big>_{\textsc{in,out}}}}$ = 
\ Event planarity ~\cite{nc4}
\item $P(n_{ch})$ = 
\ Charged particle multiplicity distribution ~\cite{nc15} \\[-3mm]
\item $N^o$ of propagating quarks vs \lead ~\cite{nc10}
\end{enumerate}

Some studies regarding DIS processes were also made, but only using DIS variables to
plot \pp data, being impossible to re-analyse DIS data in terms of \eht\! 
\cite{nc11,nc14}.

They showed no more that some agreement between the two curves but were almost
useless to decide whether \eht worked in DIS as well as it did with other
processes, as the variable used was $W^2$ that does not take into account the
\effl.

In 1984 the ISR closed and no further  intensive experimental work was planned in this
important field of non-perturbative QCD. The last relevant result was obtained in 1997 at
LEP (CERN), when the meson $\eta'$, very  rich in gluon content, was seen to be produced
in gluon induced jets as a leading particle, that is, having an anomalous high
longitudinal momentum \cite{etalead}. This completed the series of experiments aimed
to establish the universality of the \effl\  in \pp and \ee processes but nothing of
conclusive had been obtained in relation with DIS processes.

As invariant representation of \eht was chosen \qht\:
\cite{creation,nc15,nc11,usomx} in consistency with the previous use of \rads:
again a total invariant mass had been chosen to be the fundamental quantity
from which the hadronization should have been depending. Indeed \mx reduces to
\eht when evaluated in the CMS and, as all the experiments were made in
balanced colliders (where {\sc cms=lab}), then
\mbox{(\eht)$_{\text{CMS\!}}=\;$(\eht)$_{\text{LAB}}$} held, so the latter was in
effect the right quantity to  be measured.

Nowadays, because of an unjustified extention of the use of \mx 
sto DIS processes,
the almost totality of the experimental works use \mx to study the behaviour of
a given quantity, and the invariant mass is generally but wrongly believed to be the
correct representation for the \eneff. 

As a result, papers about this topic often disagree when they try to establish
if, for instance, \nch has a universal behaviour.
See, as an example, \cite{zc67, epjc11},
where the charged particle multiplicity is analysed. They both agree with the
hypotesis of the universality features, also referred to as ``fragmentation
universality''. 

On the contrary, in a more recent work \cite{osaka}, a discrepancy of 15\% is
observed between DIS and \mbox{\ee\!,} \pp mean charged particle multiplicity.

Strangely enough, the use of \mx is going on even when a fundamental quantity
as \nch has been shown to be no more ``universal'' if also DIS data are
considered.

From the viewpoint adopted here, in \cite{osaka} is proved that
\mx cannot represent the invariant quantity effectively
available for particle production and here is a first indication
that should have long since been considered: the logical step
that brought to the introduction of the \eneff\ was to separate
the interaction in the two already mentioned sub-interactions.
Now, if we insist on associating the total invariant mass to any
given particle system, we should also make the association:\\
\vspace{-5mm}
\begin{center}
\efflc \ \ $\longleftrightarrow$ \ \qlt
\end{center}
But, as it is easily seen, an invariant-mass-type quantity cannot be considered as the
variable to be associated to any sub-system into which the whole system is being divided
as
\begin{equation}
\sqrt{s} \; \neq \; M_ {\text{X}}\; + \; M_{\text{L}} 
\end{equation}

So we have to make our choice: either we separate the final state particles into
leading and hadronic, or we use invariant masses, but not both.

\section{Introduction of {\boldmath \atot}}
Firstly, let us recall that in the ``leading approach'', the
interaction is divided into two processes:
\begin{tabbing} 
Leading Effect ~~\= $\equiv$~~ \=the quantum number conservation mechanism \\[1.5mm]
Hadronization      \>$\equiv$ \>the mechanism that transforms some available quantity \\ 
                                \>~~ \>          into particles masses 
\end{tabbing}

The quantum numbers of the incident particles flow from the initial
to the final state thanks to the leading particles (usually but not
always identifiable with the so-called remnant) \cite{nc10} and
actually carried by them. Because of this ``enhanced'' dependence
from the incident particles, the leading particles usually acquire an
anomalous large longitudinal momentum: this is the Leading Effect and
these particles are said to be produced as leading. 

Then comes the hadronization, that converts what is left into the
whole of the other particles produced in the interaction. The leading
particles and their 4-momenta are no more considered in the sense
that they are not viewed as part of the hadronic system produced.
What is left after this subtraction is what is eventually studied and
analysed. This is the ``\eneff\ approach'' presented in few words.

And here is the issue: 
is it after all correct to ignore the whole leading particle as it has been
done so far in all the works published in this field? 
The answer is: absolutely not. 
In fact the leading particle leaves the vertex following a trajectory that is
not perfectly aligned to that of the incident particle. For example, if the
leading particle is the proton remnant, it is indeed true that it will have a
large longitudinal momentum, that is, it will be strongly aligned with the
incident particle axis, but not exactly nor completely.

\begin{figure}[!h] 
\begin{center}
\includegraphics[width=0.5\textwidth]{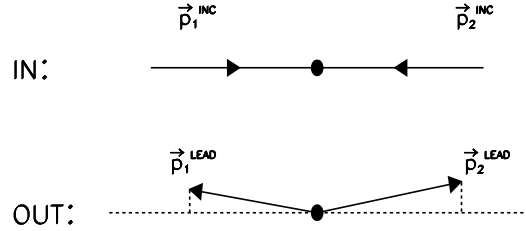}
\caption{Simplified sketch of a collision: in the initial state (top)
are represented the two incident particles, while in the final state
(bottom) only the two leading particles are shown. The transverse
contribution to their 4-momenta cannot originate from the 4-momenta 
of the incident quarks and must be a consequence of the collision,
that is, of the quark(s) loss/exchange: for this reason it cannot be
associated to QNF and must not be subtracted to \rads (see text).}
\label{collis} 
\end{center}
\end{figure}

Keeping in mind that there are some exceptions to the situation we
are about describe, let us try to show the problem by schematizing
the incident and the leading particles in a generic collision, see
Figure \ref{collis}. 

In the ``out'' sketch it is evident that the transverse component of
the leading momenta cannot originate from the incident momenta in a
direct way. In other words, the transverse leading momenta are not
those of the spectator partons (as implicitly assumed in previous
works on the \effl) but represent the partons exchanged in the
interaction.

This remark is three-dimensional, but it is straightforward to
introduce the appropriate four-dimensional version by simply
extending the formalism, that is, decomponing the energy into
its longitudinal and transverse component.

As our starting point has been the QNF, and as these quantum numbers
ought to be carried by the leading particle, it is correct to refer
to the ``motion'' quantum numbers by visualizing the trajectory of
the leading particles.

Now, if the longitudinal 4-momentum of the leading particle can be
safely associated with that of the quantum numbers propagating from
the initial state, this cannot be the same for the transverse
component. The latter must come from partons exchange with the other
incident particle, when the leading particle is the same as the
incident particle, or from the partons loss/exchange, in the case
where the leading particle is different from the incident one.

\noindent
This simple ``geometrical'' remark immediately suggests that:
\begin{quote} \vspace{-1mm}
The component of the $q^{\textsc{lead}}_{\textsc{tot}}$ that is transverse to the QNF 
is not leading. As such, it must not be subtracted from \rads\ to get the Total
\aqs.%
\footnote{Here a short note about the name chosen for this new
invariant is appropriate: it is simply the most compact among the
correct names. In fact, it would be wrong to use the word ``energy'',
as well as it is not correct to use the term ``mass''. In fact \atot is not
an invariant-mass-like quantity. So, the good practice to call things
with a name that resembles their nature has been applied.}
\end{quote}

The longitudinal component of $q^{\textsc{lead}}_{\textsc{tot}}$ will be called
\ltot and represents the ``cost'' to be paid in a given event to force the quantum
numbers conservation. What is left from \rads\,is the invariant variable to be
associated with the multihadronic final state and that represents the ``slice''
of \rads effectively available to its production.  In other words, it is the
Lorentz-invariant quantity that should be used to analyse and describe
every experimental result. Summarizing we have:
\begin{equation*}
\sqrt{s} \ \ \longrightarrow \ \ 
\begin{cases}
\;\mathcal{L}_{\text{TOT}} &\text{goes to the \effl}\\[3mm]
\;A_{\text{TOT}} &\text{goes to the hadronization}
\end{cases} 
\end{equation*}

The indices ``\te{tot}'' stays for ``\te{total}'' and reminds us that
the whole product of the interaction is being considered. Anyway, in
principle, nothing would prevent us from considering one hemisphere
only by defining $A_1$ for hemisphere 1, and $A_2$ for hemisphere 2. 
It is easy, in such case, to change the definition of \atot given
below: it suffices to consider the leading (or hadronic)
quadrimomentum measured in one hemisphere instead that looking at
both (see next section for a consistent definition given in equation
\ref{uno}). This means that it is possible to use the variables
introduced later in the paper to study target experiments too.
Useless to say that this agrees, and is imposed by, the universality
of the \effl. This situation will not be analysed here and when we
talk about the \aqs\ we are actually talking about the total \aqs.

It is important at this point to distinguish two kinds of \effl\ that
will be called here direct and indirect. The first type is observed
in all the collisions but the annihilation processes. The second one
regards the annihilation processes only.

But how is it possible to have a \effl\ if the incident particles
annihilate?  Actually, it is still possible to talk about a \effl\
but it relates no more to the propagation of the quantum numbers of
the incident particles. Rather, it is referred to the quantum numbers
of the particles formed by pair production from the $\gamma/Z^0$
produced at the annihilation vertex.

This second type of \effl\ has been observed in \ee annihilations (it
is present at a 1\% rate) and the most important studies regarded the
production of the charmed meson $D^*$ \cite{creation} and that of the
``gluonic'' meson $\eta'$ \cite{etalead}.

As for the $D^*$, the propagating quantum numbers are those of a $c$
or $\bar{c}$ quark while, when the $\eta'$ is concerned, its strong
gluonic component tells us that it is carrying the quantum numbers of
the gluon: in fact, the $\eta'$ appears to be leading only when
emerging from a gluon induced jet (3 or 4-jet events).

Why this distinction? Apart from the conceptual, there is a practical
reason to insist on it. In fact, the calculations performed later
cannot be referred to the indirect \effl,  and the given definition
of \atot itself ought to be changed. It would not be a difficult
task, and the appropriate formulation in this case is also given
below, but the experimental evidence has not been pursued yet.

It is evident anyway that, even in such situation, some projection
must be done. We can guess that the axis on which $q^{\textsc{lead}}$
should be projected is the axis given by the mean momentum of the jet
containing the leading particle, namely,  the ``charmed'' jet or the
``gluonic'' jet respectively. 

Keeping in mind this distinction, it is now possible to find a
mathematical expression for the scalar quantity available to the
energy-into-mass transformation.

\section{Formal definition of {\boldmath \ltot and \atot}}
Following the remarks made in the previous section and
using the minkowskyan metric
instead of the euclidean, it is immediate to write the expression for \ltot
\begin{equation}
\mathcal{L}_{\text{TOT}} \:\equiv  \: q^{\te{lead}}_{\te{tot}}\cdot
\frac{q^{\te{inc}}_{\te{tot}}}{\sqrt{s}} 
\end{equation}

According to our point of view, this represents the cost of the quantum numbers
conservation in a given event.
What is left is what the physics assign to particle production
\begin{equation}
A_{\text{TOT}} \ \equiv \ \sqrt{(q^{\te{inc}}_{\te{tot}})^2}-
q^{\te{lead}}_{\te{tot}}\cdot 
\frac{q^{\te{inc}}_{\te{tot}}}{\sqrt{s}}
\label{eq:at}
\end{equation}
or, in compact notation
\begin{equation}
A_{\text{TOT}} \equiv \sqrt{s}- \mathcal{L}_{\text{TOT}}
\label{eq:comp}
\end{equation}
This is the correct formal definition of \atot but we should now find 
a manageable expression to be used below. 
A possible one is given by writing
\begin{equation}
\sqrt{(q^{\te{inc}}_{\te{tot}})^2}=\frac{(q^{\te{inc}}_{\te{tot}})^2}
{\sqrt{(q^{\te{inc}}_{\te{tot}})^2}}
\end{equation}
and substituting in (\ref{eq:at}) to get
\begin{equation}
A_{\text{TOT}}=q^{\te{had}}_{\te{tot}}\cdot \frac{q^{\te{inc}}_{\te{tot}}}{\sqrt{s}}
\end{equation}
that is, \atot is the projection of the sum of all the 4-momenta of the produced hadrons
(excluding of course the leading particles) on the axis given by the incident particles.

Incidentally, this last remark suggests the only possible consistent definition for $A_1$,
that, as discussed in the previous section,
is the variable to be used when considering a single hemisphere (in which case
also $A_2$ could be of interest) or a target experiment:
\begin{equation}
A_1=q^{\te{had}}_1\cdot \frac{q^{\te{inc}}_{\te{tot}}}{\sqrt{s}}
\label{uno}
\end{equation}

It must be recalled again that this result is only valid for the direct \effl. When we
are concerned with the indirect \effl\ we should write
\begin{equation}
\mathcal{L}_{\text{TOT}}(\textsf{ind}) \:\equiv  \: q^{\te{lead}}_{\te{tot}}\cdot
\frac{q^{\te{jet}}_{\te{tot}}}{\sqrt{(q^{\te{jet}}_{\te{tot}})^2}} 
\end{equation}
and coherently modify the expression for \atot\!. The last is valid provided that there is
no further \effl\ observed in the other jets (in which case the definition of \ltot
should be changed accordingly).

First of all, the consistency with previous works must be tested. These all
clearly showed that, if the variable 
\begin{equation}
E^{\textsc{had}}_{\textsc{tot}} =
E^{\textsc{inc}}_{\textsc{tot}} - E^{\textsc{lead}}_{\textsc{tot}} 
\end{equation}
is used to perform data analysis, then universality features manifest
themselves: a whole set of measured quantities shows the same behaviour
independently of everything but \mbox{\eht\!.} So the  value of \atot in terms
of \eht must be estimated.

The colliders where the \effl\ was studied (ISR, LEP, HERA$(e^+e^-)$) were all
balanced, that is, the incident particles had the same energy. In this case
\begin{equation}
A_{\text{TOT}} = (E^{\te{had}}_{\te{tot}})_{\te{lab}}= 
(E^{\te{had}}_{\te{tot}})_{\cms}~~~~~\text{(balanced colliders)}
\end{equation}
must hold. In fact, what we get
if we specialize to the CMS is:
\begin{equation} 
\begin{split}
A_{\text{TOT}} \ & =\  \:q^{\te{had}}_{\te{tot}}\cdot
\frac{q^{\te{inc}}_{\te{tot}}}{\sqrt{s}}\: =\:
(E^{\te{had}}_{\te{tot}};\vec{p}\,^{\te{had}}_{\te{tot}})_{\te{cms}}\cdot 
\frac{(E^{\te{inc}}_{\te{tot}};\vec{0})_{\te{cms}}}{\sqrt{s}} \ =\\[4mm]  & =\ \bigg( 
\frac{E^{\te{had}}_{\te{tot}}\cdot E^{\te{inc}}_{\te{tot}}}{E^{\te{inc}}_{\te{tot}}}
\bigg)_{\te{cms}} = \ (E^{\te{had}}_{\te{tot}})_{\te{cms}}
\label{eq:aeht}
\end{split} 
\end{equation}
and there are no consistency troubles as we recover the variable used
in that early works performed in the CMS limit. The previous equation
also shows that:
\begin{equation}
A_{\text{TOT}} \neq (E^{\textsc{had}}_{\textsc{tot}})_{\te{non cms}}
\end{equation}
that is a very important result and tells us that, in general, \emph{\eht is
not the \eneff}. In other words, at an unbalanced collider, \emph{the direct
use of \eht will not work and is not correct}.

This means that the universality features were discovered thanks to
the fact that the experiments were incidentally conducted in the
system where the \aqs\ matches the energy of the hadrons produced
(excluding the leading particles). Probably, this is also the reason
that induced the discoverers to think that a non-invariant quantity
like the hadronic energy, could have had such a fundamental role. Had
those colliders been unbalanced (as is HERA nowadays), the discovery
of the universality features could not have been made.

In order to be exhaustive, it should be said that some late works
made at the ISR were performed using \mx instead of \eht
\cite{creation,nc15,nc11,usomx} but the Universality Features were
evident anyway, even if, in general, \mx$\neq$ \eht\!. How is it
possible? In chapter 8, it will be showed how the reason for \mx
worked so well in those occasions, lies in the particular cuts made
on the data set at that time as well as in the low energies and
resolutions.

\section{Relation between {\boldmath \mx and \atot}}
As mentioned, in general, \atot$\neq$ \mx\!. The best way to get a useful
relation between these two variables is to estimate the quantity
$(A_{\text{TOT}}^2-M_{\text{X}}^2)$:
\begin{align}
&A^2_{\text{TOT}}=(\sqrt{s}-q^{\te{lead}}_{\te{tot}}\cdot \frac{q^{\te{inc}}_{\te{tot}}}
{\sqrt{s}})^2 = s-2q^{\te{lead}}_{\te{tot}}\,q^{\te{inc}}_{\te{tot}}+
\frac{(q^{\te{lead}}_{\te{tot}}\,q^{\te{inc}}_{\te{tot}})^2}{s}
\label{eq:a}\\[4mm]
&M^2_{\text{X}}=(q^{\te{inc}}_{\te{tot}}-q^{\te{lead}}_{\te{tot}})^2= 
s-2q^{\te{inc}}_{\te{tot}}\,q^{\te{lead}}_{\te{tot}}+(q^{\te{lead}}_{\te{tot}})^2 
\label{eq:m}
\end{align} 
so that the difference is all in the last terms:
\begin{equation}
(A_{\text{TOT}}^2-M_{\text{X}}^2)=
\Bigg[ \frac{(q^{\te{lead}}_{\te{tot}}q^{\te{inc}}_{\te{tot}})^2}{s} - 
(q^{\te{lead}}_{\te{tot}})^2 \Bigg]
\end{equation}

This is the correct invariant expression for the difference we
are estimating, but we also  need an expression that is function
of some measured quantities in order to perform the calculations
that follow. 

To simplify the notation we introduce the following shortcuts 
for the indices Total, Incident, Hadronic and Leading:
\begin{eqnarray*} \vspace*{-6mm}
\te{inc} &\longleftrightarrow &\te{i}\\[-2mm]
\te{lead} &\longleftrightarrow &\te{l}\\[-2mm]
\te{had} &\longleftrightarrow &\te{h}\\[-2mm]
\te{tot} &\longleftrightarrow &\te{t}
\end{eqnarray*}
Calculating in the CMS we get
\begin{equation} 
\begin{split}
\Big[ A^2_{\text{TOT}}-M^2_{\text{X}}  \Big] &=\bigg[ \frac{(q^L_Tq^I_T)^2}{s}-(q^L_T)^2 \bigg]
=(E^L_T)^2_{\cms}-(q^L_T)^2=\\[4mm]
&=(E^L_T)^2_{\cms}-[(E^L_T)^2_{\cms}%
-(\vec{p}\,^L_T)^2_{\cms}]=(\vec{p}\,^L_T)^2_{\cms}=(\vec{p}\,^H_T)^2_{\cms}\\[-2mm] 
\label{eq:am}
\end{split} 
\end{equation}
This also follows by observing that
\begin{equation}
\begin{cases}
A^2_{\text{TOT}}=(E^H_T)_{\cms}^2\\[2mm]
M^2_{\text{X}}=(E^H_T)_{\cms}^2 - (\vec{p}\,^H_T)_{\cms}^2
\end{cases}
\end{equation}
so that
\begin{equation}
\Big[ A^2_{\text{TOT}}-M^2_{\text{X}}  \Big] \ = \ (\vec{p}\,^H_T)_{\cms}^2
\tag{\ref{eq:am}\,$'$}
\label{eq:amam}
\end{equation}
\vspace{2mm}

\noindent The importance of equation (\ref{eq:amam}) is
threefold:%
\begin{enumerate}
\item It shows that \atot $\ge$ \mx that is, the hadronic system
``has more 4-scalar at disposal'' for particle production than
what was believed so far. This implies that more  particles can
be produced than the value of Mx could suggest (and at constant
Mx), depending  on the total 3-momentum of the leading
particle(s) (see point 3 below) as evaluated in the CMS;
\item It gives a very simple recipe to re-analyse data that were
wrongly plotted vs \mx\!\!: it is in fact sufficient to
substitute (event by event) \mx with the sum in quadrature of
\mx and  $(\vec{p}\,^L_T)^2_{\cms}$:

\begin{align}
&M_{\text{X}} \longrightarrow \sqrt{M^2_{\text{X}}+(\vec{p}\,^L_T)^2_{\cms}}
~~~~~~~~~~~~~~~\text{or} \label{eq:unom}\\[4mm]
&A_{\text{TOT}}\;=\;\sqrt{M^2_{\text{X}}+(\vec{p}\,^L_T)^2_{\cms}} 
\tag{\ref{eq:unom}\,$'$}
\label{eq:amxpl}
\end{align}
Here the knowledge of $(\vec{p}\,^L_T)^2_{\cms}$ is necessary,
otherwise the previous formula becomes useless and another
method, to be showed later in the paper, must be employed; \item
It highlights the importance of the unbalance in
$(\vec{p}\,^L_T)^2_{\cms}$ between the two hemispheres: what
makes the difference between \atot and \mx has nothing to do
with the absolute value of the \effl\ but depends only on its
unbalance: the more this unbalance is, the more Mx differs from
Atot, the less \mx is able to
highlight the universality features. In other words, you can
have a strong, a medium, or a small \effl: if the momenta of the
two leading particles detected are the same, then you will
always find that \atot= \mx\!, regardless of their absolute
value.
\begin{equation}
A_{\text{TOT}}= M_{\text{X}}= (E^{H}_{T})_{\cms} \hspace{1.2cm} 
\big( \text{ if~~~} \ (\vec{p}\,^H_1+\vec{p}\,^H_2)_{\cms}=0\ \big)
\end{equation}
\end{enumerate}

In view of the application of \atot to compare \pp and \ee data, another
fact must be discussed here, not only for future reference, but also to check the
compatibility with previous results.

When the confrontation between \pp and \ee data were made, the latter were
plotted in terms of
\rads\!. When we do not have a \effl\ in \ee annihilation, the relation \rads=\mx holds. 
But, as stated before, a strong \effl\ in \ee processes is only present at a 1\% rate. 
In other events the \effl\ can be neglected.

Besides, when an average among all the events is performed, the 1\% showing a large
\effl\ is easily overcome by the remaining 99\% and becomes undetectable,
in the sense that the difference between
\mx and \rads is far below the experimental uncertainties. This means that, for those
samples of \ee data, you have 
\[ \sqrt{s} = M_{\text{X}} \]
and, as \ee colliders always have the feature {\sc lab=cms}, the relation 
\[  M_{\text{X}} = E^{\textsc{had}}_{\textsc{tot}} \]
holds too.
\noindent In other words, for our purposes, in \ee annihilations:
\begin{equation}
A_{\text{TOT}}= M_{\text{X}}= \sqrt{s}= E^{\textsc{had}}_{\textsc{tot}}
\text{~~~~~~~~~~~~~~~~(\ee ~annihilations)}
\end{equation}
and we have not to worry about which variable was used. This will turn out to be
important later on, where the experimental evidence of what has been
conjectured so far is provided.

Before considering this evidence, I would like to address what I consider 
a very important conceptual issue. Conservation laws at vertices are
always expressed in terms of 4-vectorial quantities (like 4-momenta, tensors,
etc). This is true even in the \effl\ framework, where the conservation law
reads:
\begin{equation}
q^{\te{had}}_{\te{tot}} = q^{\te{inc}}_{\te{tot}} - q^{\te{lead}}_{\te{tot}}
\label{eq:qmc} 
\end{equation}

On the contrary, when the experimental results are analysed, you always have to do with
some scalar quantity like \rads or \mx (as the final product of any
analysis usually is some scalar function of one variable).

The approach pursued here provides a consistent jump between the former and the latter as
it allows to get from (\ref{eq:qmc}) the following scalar conservation law:
\begin{equation}
\sqrt{s} = A_{\text{TOT}} + \mathcal{L}_{\text{TOT}}
\tag{\ref{eq:comp}\,$'$}
\end{equation}

It seems to be a trivial step but, as previously noted, the
world-wide use of \mx shows as it often happens that we do not
recognize simple facts as a conservation law violation: the very use
of \mx in the framework of the \eneff\ violates the 4-momentum
conservation law at interaction vertices and, consequently, of the
whole interaction itself.

\section{Evidence of {\boldmath \nch vs \atot universality}}
In this section we show  how it is possible to recover QCD
universality by using \atot to plot data coming from different
experiments. The proof is obtained by rescaling a result given in
another paper \cite{osaka}. \footnote{Presently I have no access to
row data, so I could not make a direct analysis.}

\begin{figure}[!h]
\begin{center}
\hspace*{-7mm} 
\includegraphics[width=0.65\textwidth]{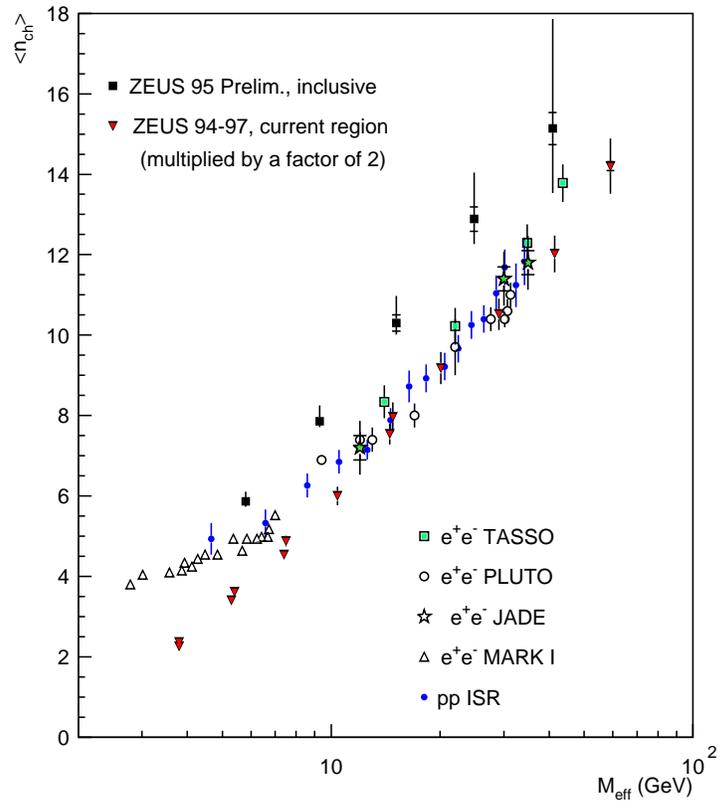}
\caption{Figure taken from \cite{osaka} showing the mean charged particle
multiplicity in DIS events plotted as a function of  \meff and confronted with
the same quantity measured for other processes. The inner bars show the
statistical errors, the outer ones show the statistical and  systematic errors
added in quadrature.}
\label{osaka}
\end{center}
\end{figure}

It is appropriate here to briefly summarize the results of that work.
It is a study of the quantity \nch in DIS processes. The main result
is that \nch\!$_{\!\!\te{dis}}$ is more than 15\% higher than \nch
measured for \pp and \ee processes. This is well above the
experimental uncertainties, so the conclusion was: DIS processes
disagree with the  \effl\ phenomenology and bring to a QCD
non-universality. Figure \ref{osaka} shows  the variable \nch as
measured in \cite{osaka}.

\begin{figure}[!h]
\begin{center}
\includegraphics[width=0.5\textwidth]{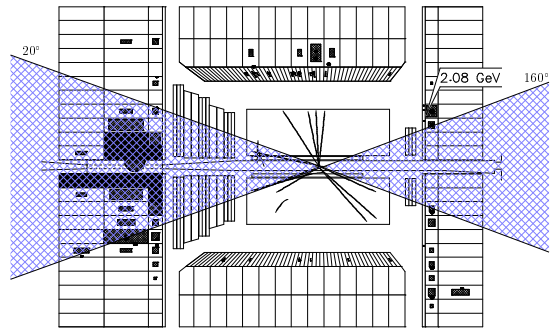}
\caption{Typical DIS event in the ZEUS detector. The polar angle cut
is shown: shaded areas are not considered in the analyses. 
Figure taken from \cite{osaka}.}
\label{intpol}
\end{center}
\end{figure}

The study was conducted in a reduced phase space. The charged tracks and the
energies were only measured in the polar angle interval $\theta_{\te{lab}}\!
\in[20^0;160^0]$,  see Figure \ref{intpol}. 

This cut is justified by a large improvement in the experimental
resolutions and, as proved in the same paper by means of MC
simulations, it does not affect the result in any way. The other
cuts are not relevant and are those normally used to obtain an almost
pure set of DIS events. More details may be found in \cite{osaka} or
below.

Because of this cut in the polar angle, the variable \mx is
substituted with \mbox{\meff\!:} it is again a hadronic invariant
mass but this time it is not the total one; in fact, it is that
measured in the $\Delta \theta$ considered. Again, MC simulations
proved that the dependence between \nch and \mx \mbox{(\meff\!\!)} is not
affected by the phase space reduction.

From now on, unless otherwise stated, all variables refer to the
reduced phase space without introducing any new notation to recall
this fact. The only variable which notation will be changed is \mx to
\meff in order to help any comparison with \cite{osaka}.

Of course, the correction factor is the ratio between \atot and \mx\!
and will be denoted by $a$. This factor will be applied to the abscissa
in the plot showed in Figure \ref{osaka}, and will be obtained in a
high energy approximation: this will cause a very slight mismatch at
low energies even when \atot is used for re-plotting.  This
approximation could be avoided by re-analysing the data in terms of
\mbox{\atot\!\!.}

To give an estimation of the correction factor $a$ that is suitable
for our purposes, the general result (\ref{eq:am}) is useless as
$\vec{p}^{\te{\:lead}}_{\:\te{tot}}$ was not measured. Also that
result is strictly valid only in the full phase space. Thus the
correction has been estimated by averaging over the two variables
on which the said ratio depends, namely $\theta$ and $y$ (as showed
below). The latter is a new variable that is defined in the next
section and that expresses the unbalance in hadronic energy between
the two hemispheres in a way that is suitable for our calculation.

\noindent We proceed by estimatimating the quantity 
$s \cdot (A_{\text{TOT}}^2-M^2_{\text{eff}})$: \\[2mm]
\begin{equation}
\begin{split}
s\cdot[A_{\text{TOT}}^2-M_{\text{eff}}^2]
&=[E^HE^I-\vec{p}\,^H\vec{p}\,^I]^2-[(E^I)^2-(\vec{p}\,^I)^2][(E^H)^2-(\vec{p}\,^H)^2]=\\[4mm]
&=(E^HE^I)^2+(p^Hp^I\cos \theta)^2-2E^HE^Ip^Hp^I\cos \theta-\\[2mm]
&\qquad-(E^HE^I)^2-(p^Ip^H)^2+(E^Ip^H)^2+(E^Hp^I)^2=\\[4mm]
&=(p^Hp^I)^2(\cos^2\theta-1)-2(E^HE^Ip^Hp^I)\cos \theta+\\[2mm]
&\qquad+(E^Ip^H)^2+(E^Hp^I)^2=\ldots  \label{eq:diff}  
\end{split} \vspace*{-3mm}
\end{equation} 
where the index \te{tot} has been suppressed and the absolute value of 3-vectors 
is denoted by suppressing the arrow above the vectors. 

Of course $\theta$ is the angle between $\vec{p}\,^I$ and $\vec{p}\,^H$ that is, 
considering HERA kinematics, between $\vec{p}\,^H$ and the $z$ axis:
\begin{equation}
\theta =  \angle \:(\vec{p}\,^H_T\, ; \, \hat{z})
\end{equation}
This means that $\theta$ is the polar angle, and will be identified with it 
from now on. Introducing the high energies approximation:
\begin{equation*}
\begin{cases} 
p^H\!\simeq E^H\\  p^{\,I}\simeq E^I
\end{cases}
\end{equation*}
and substituting in (\ref{eq:diff}) we get:
\begin{equation}
\begin{split}
\ldots&=(E^HE^I)^2(\cos^2 \theta-1)-2(E^HE^I)^2\cos \theta+2(E^HE^I)^2=\\[4mm]
&=(E^HE^I)^2(1-2\cos \theta+\cos^2\theta)=\\[4mm]
&=[E^HE^I(1-\cos \theta)]^2 
\end{split} \tag{\ref{eq:diff}\,$'$}
\end{equation}

\noindent To sum up:
\begin{equation}
s\cdot[A_{TOT}^2-M_{\textrm{eff}}^2]\ \simeq\  [E^HE^I(1-\cos \theta)]^2\ 
\ \ \ \ \quad (E^{H,I} \simeq p^{H,I})
\end{equation}

In order to further simplify the last expression we are now forced to choose a
reference frame. Again the best choice is the CMS, both to simplify $s$ and for
sake of visualization. Besides, it is in general the most ``balanced'' system,
and this helps in avoiding as many troubles as possible with the high energies
approximation. The only complication induced by this choice is that it forces
to perform the needed transformations on data collected in the \te{lab} system.

\noindent Calculating in the CMS we obtain:
\begin{equation}
s\cdot[A_{\text{TOT}}^2-M_{\text{eff}}^2]\ \simeq\ 
(E^H)^2_{\cms}(E^I)^2_{\cms}(1-\cos \theta)^2_{\cms}
\end{equation}
and using 
\begin{equation}
s=(E^{\te{inc}}_{\te{tot}})^2_{\cms}
\end{equation}
it becomes
\begin{equation}
[A_{\text{TOT}}^2-M_{\text{eff}}^2]\simeq (E^H)^2_{\cms}(1-\cos \theta)^2_{\cms}
\label{eq:atmxosa}
\end{equation}

As already mentioned, the ratio between \atot and \mx is a
function of two variables: $\Delta \theta$ and $y$. This means that, after having
imposed the cuts used in the paper to be corrected, considering the collision parameter
at HERA, and finally averaging over these two variables, we are left with a
linear dependence of \atot from \meff\!:
\begin{equation}
A_{\text{TOT}}\ \propto\ M_{\text{eff}} \qquad \text{(fixed kinematical conditions)} 
\end{equation}

\noindent
Thus, if in general we have
\begin{equation}
\frac{A_{\text{TOT}}}{M_{\text{eff}}} \equiv a(\Delta \theta , y) \label{eq:aa}
\end{equation}
the previous remark allows the calculation of the constant 
$a=\big< a(\Delta \theta , y) \big>_{(\Delta \theta , y)}$ 
that expresses the ratio between \atot and \mx at fixed kinematical conditions:
\begin{equation}
a \; \equiv \; \big< a(\Delta \theta , y) \big>_{(\Delta \theta , y)} \qquad 
\text{(fixed kinematical conditions)}
\end{equation}
Note that $a$ is invariant being the ratio between two invariants.

The dependence of $a$ from $\Delta \theta$ and $y$ will be used later
to show how it is possible to obtain a different value for $a$ by
tuning these parameters. This fact supports the correctness of the
introduction of \atot\!, as we obtain a value that allows DIS data to
match with other processes curves if and only if the cuts used in
\cite{osaka} are taken into account.

\noindent Using the notation introduced in (\ref{eq:aa}), equation (\ref{eq:atmxosa}) can be
rewritten as:
\begin{equation}
[a^2(\Delta \theta , y)-1] \cdot M^2_{\textrm{eff}} \; \simeq \; 
(E^H)^2_{\cms}(1-\cos \theta)^2_{\cms}
\tag{\ref{eq:atmxosa}\,$'$}
\end{equation}
or, isolating $a$,
\begin{equation}
a^2(\Delta \theta , y) \; \simeq \; 
1+\frac{(E^H)^2_{\cms}}{M^2_{\textrm{eff}}}\cdot(1-\cos \theta)^2_{\cms}
\label{eq:ageneral}
\end{equation}

\noindent The relevant cuts and numerical values are:
\begin{itemize}
\item  $\theta_{\textsc{lab}}\in [20^0;160^0]$: the already mentioned polar angle cut
\cite{osaka}. 
It must be re-evaluated in the \cms;
\item  $(E'_e)^{\textsc{lab}}_{\textsc{min}}=8\ \text{Gev}$,
where the prime is used to refer to
the final state and the ``$e\,$'' indicates the final state positron (this cut is
imposed to guarantee an almost pure level of DIS events). Even this value  must be
re-evaluated in the \cms;
\item $\beta_{\textsc{boost}}$\,=\,0.935 is the value of the boost between the \te{lab}
and the \cms\ at HERA.
\end{itemize}
and we are now ready to estimate the correction factor we are seeking,
that is, the mean of $a$ with respect to the variables it depends on:
\begin{equation}
a \; \equiv \; \sqrt{1+\frac{\big< (1-\cos \theta)^2_{\cms} \big>\phantom{2}}%
{\big< M_{\textrm{eff}}/(E^H_T)_{\cms} \big>^2}}
\label{eq:val}
\end{equation}

\subsection{Average over the polar angle}
In this subsection only a prime is used to indicate \cms\ variables 
while unprimed varibles refer to the \te{lab} system.
Writing the appropriate transformations for the 4-momenta
\begin{equation}
\begin{cases}
E'=\gamma(E-\beta p_z)\\[2mm]
p'_z=\gamma(p_z-\beta E)
\end{cases}
\end{equation}
and using again the high energies approximation in the form
\begin{equation}
\begin{cases}
p_z=p \cos \theta \simeq E \cos \theta\\[2mm]
p'_z=p' \cos \theta' \simeq E' \cos \theta'
\end{cases}
\end{equation}
we get
\begin{equation}
\begin{cases}
E'=\gamma(E-\beta E \cos \theta)\\[2mm]
E' \cos \theta'=\gamma(E \cos \theta-\beta E)
\end{cases}
\end{equation}
so that
\begin{equation}
\cos \theta' \simeq \frac{\gamma E(\cos \theta-\beta)}{\gamma E(1- \beta \cos \theta)}=
 \frac{\cos \theta-\beta}{1- \beta \cos \theta}
\end{equation}
that is the well known formula for the aberration of light. The use of the last result 
is again cause of some overestimation of $a$ at low energies.

Working in the \cms\ allows to perform the mean over the polar angle by simply averaging
over the distribution obtained in (\ref{eq:ageneral}) as, if the sample is large enough, 
$\big< \theta \big>_{\cms} \simeq \, \pi/2$\,.

\noindent
Using the interval given in \cite{osaka}, \!$\theta \in[20^0;160^0]$, and performing the
needed transformations we get
\begin{equation}
\begin{cases}
20^0 &\longrightarrow ~87.895^0~ ~\:=~ 1.536~ \text{rad}\\[2mm]
160^0 &\longrightarrow ~176.305^0~ =~ 3.077~ \text{rad.}
\end{cases}
\end{equation}
These are the extremes of integration to be used. The integral yields
\begin{equation}
\big< (1-\cos \theta)^2 \big>_{\cms} \: =\: \frac{{\displaystyle \int_{1.536}^{3.077} 
(1-\cos \theta)^2 \ud \theta}}{\Delta\theta} \: \simeq \: 2.68
\end{equation}
so that
\begin{equation}
a(y) \: \equiv\:  \big<\, a(\Delta \theta,y)\, \big>_{(\Delta \theta)} \: \simeq \:
\sqrt{1+\frac{2.68}{(M_{\text{eff}}/E^H)^2}}
\end{equation}

\subsection{Average over the new unbalance variable ``$y$''}
As anticipated above, we now define a variable that suitably 
represents the unbalance in hadronic energy between the two hemispheres. 
We call this variable $y$ and define it as follows:
\begin{equation} \begin{cases} 
E^H_{1,p}=y\cdot E^H_T\\[2mm] E^H_{2,e}=(1-y)\cdot E^H_T
\label{eq:ydef}
\end{cases} \end{equation}
that is, $y$ is the fraction of hadronic energy measured in hemisphere 1 (that of the
``target region'', where the proton remnant is detected).

Using $y$ we can express \mx as a function of \eht in a very useful way. In fact,
assuming that $\hat{k}\,^H_1 \simeq -\hat{k}\,^H_2$ (where $\hat{k}\,^H$ represents the
hadronic 3-momentum versor) we have
\begin{equation}
\begin{split}
M_{\text{X}} \equiv \sqrt{(q_{\te{tot}}^{\te{had}})^2} &= 
\big[ (q^H_1)^2 + (q^H_2)^2 + 2E^H_1E^H_2 -%
2\vec{p}\,^H_1 \cdot \vec{p}\,^H_2 \big]^{1/2} \simeq \\[2mm]
& \simeq \big[ 2E^H_1E^H_2 + 2E^H_1E^H_2 \big]^{1/2} = 2 \sqrt{E^H_1E^H_2}
\end{split}
\end{equation}
that together with definition (\ref{eq:ydef}) yields
\begin{equation}
M_{\text{X}} \simeq 2 \sqrt{E^H_1E^H_2} = 2E^{\te{had}}_{\te{tot}} \sqrt{y(1-y)}
\label{eq:mxdiy}
\end{equation}

We remark that this rescaling method is applicable only if we have no
more than 2 hadronic jets. In case of 3 or more jets, another method should be
used  or a direct data analysis must be performed.

The hadronic energies we are considering are ``real'' hadronic energies, that
is, the energies effectively produced at the interaction
vertex, regardless of the cuts applied to perform the analysis.

This does not create problems in case of a phase space reduction as there is no
dependence of \mx(\eht\!\!) from the phase space \cite{osaka}. On the contrary, the cut
in the final positron energy must be considered, as the events that do not meet this
condition do not become part of the statistics. 

As $\big< y \big>_{\cms\,} = 1/2$\, holds, for the same reason exposed in the previous
section, it is sufficient to calculate the mean of the distribution obtained in the
(\ref{eq:mxdiy}). Had there been no cuts in the leading particles energy, this integral
would have yielded $\pi/8$, that is half the area of a circle of radius 1/2. This is not
our case and we will have to take into account the relevant cuts, one of which has been
already mentioned and is:
\[ (E'_e)^{\textsc{lab}}_{\textsc{min}} = 8\: \text{Gev} \]
that, as usual, must be re-evaluated in the \cms :
\begin{equation}
\begin{split}
(E'_e)^{\textsc{cms}}_{\textsc{min}} \: &=\: \gamma \big[
(E'_e)^{\textsc{lab}}_{\textsc{min}}- \vec{\beta}\cdot \vec{p}\,^{\textsc{lab}}_z \big]
\: \simeq \: (E'_e)^{\textsc{lab}}_{\textsc{min}} \cdot \frac{1+\beta}{\sqrt{1-\beta^2}}
\: =\\[2mm] &=(E'_e)^{\textsc{lab}}_{\textsc{min}} \cdot \sqrt{\frac{1+\beta}{1-\beta}}
\: \simeq \: 5.467\: (E'_e)^{\textsc{lab}}_{\textsc{min}} \: \simeq \: 43.7~ \text{Gev}
\end{split}
\end{equation}
where we have:
\begin{itemize}
\item considered that $\hat{\beta}$ e $\hat{z}$ are antiparallel, from where the change in
sign at the second equality
\item used the high energy approximation and the strong collinearity of the 
$e^+_{\textsc{out}}$ (again at the second equality)
\item used the value of $\beta_{\textsc{boost}}$ for $\beta$
\end{itemize}

The highest leading longitudinal momentum in the target region 
is detected when a proton takes the role of the leading particle: 
as this maximum value, we use that representing the limit between 
the leading physics and the diffractive physics 
(in the sense described in \cite{evid} or \cite{creation}), namely:
\begin{equation}
(E^{\te{lead}}_{1,p})_{\te{max}} \: = \: (x_F)_{\te{max}}^{\cms}
\cdot E^{\te{inc}}_{1,p}\: 
=\: 0.8\: E^{\te{inc}}_{1,p} \label{eq:xf}
\end{equation}
where $x_F$ is the Feynman variable that represents, in practice, 
the fractional longitudinal momentum of the leading particle.

It should be noted that the relation between energies or momenta 
expressed in fractional terms are approximately invariant. In fact:
\begin{equation}
\begin{split}
&\begin{cases}
E'_{\te{lead}} = \gamma (E^L-\vec{\beta}\cdot \vec{p}\,^L)\\[4mm]
E'_{\te{inc}} = \gamma (E^I-\vec{\beta}\cdot \vec{p}\,^I)
\end{cases} \longrightarrow\\[4mm]
& \longrightarrow ~~ 
\frac{E'_L}{E'_I} = \frac{\gamma (E^L-\vec{\beta}\cdot \vec{p}\,^L)}{\gamma (E^I-\vec{\beta}\cdot
\vec{p}\,^I)} \simeq \frac{\gamma (1-\beta)}{\gamma (1-\beta)} \cdot \frac{E^L}{E^I} =
\frac{E^L}{E^I}
\end{split}
\end{equation}
so that the energy of a particle expressed with the Feynman variable is about the same in
two different frames at high energies and in a collinear approximation.
The leading particle, as such, respects these approximations.

The two values just obtained allow to find the largest interval in hadronic energy 
compatible with the cuts:
\begin{equation*}
\begin{cases}
(E^H_1)_{\textsc{min}}^{\cms}={\displaystyle \frac{\sqrt{s}}{2}-\frac{4}{5}\cdot 
\frac{\sqrt{s}}{2}}\: \simeq\: 30.0\: \text{Gev} \\[4mm]
(E^H_2)_{\textsc{max}}^{\cms}={\displaystyle 
\frac{\sqrt{s}}{2}-(E'_e)^{\textsc{cms}}_{\textsc{min}}}\: \simeq\: 106.5\: \text{Gev}
\end{cases}
\end{equation*}
from which the minimum for $y$ immediately follows:
\begin{equation}
y_{\textsc{min}} = 
\frac{(E^H_1)_{\cms}^{\textsc{min}}}{(E^H_1)_{\cms}^{\textsc{min}}+
(E^H_2)_{\cms}^{\textsc{max}}} \simeq 0.220
\end{equation}

In order to give an estimation for $y_{\textsc{max}}$, it is
necessary to make some assumptions regarding $x_F^{\textsc{min}}$ and 
$(E'_e)^{\textsc{cms}}_{\textsc{max}}$\,. A good choice, in view
of a comparison with ISR data, is to adopt the same value of 
$x_F^{\textsc{min}}$ chosen at ISR, namely \mbox{0.4\,:}

\[ (x_F)_{\textsc{min}}^{\cms} = 0.4\: E^{\te{inc}}_{1,p} \]

It will also be assumed that the final state positron loses at least the 30\% of its 
energy, value above which the diffractive events begin to prevail over the DIS ones:
\[ (E'_e)^{\textsc{cms}}_{\textsc{max}} = 0.7 {\displaystyle \frac{\sqrt{s}}{2}}\: 
\simeq\: 105\: \text{Gev} \]
From the previous two equations we get
\begin{equation}
\begin{cases}
(E^H_1)_{\cms}^{\textsc{max}}={\displaystyle \frac{3}{5}\cdot \frac{\sqrt{s}}{2}}\:
\simeq\: 90.1\: \text{Gev} \\[4mm]
(E^H_2)_{\cms}^{\textsc{min}}={\displaystyle 
\frac{\sqrt{s}}{2}-(E'_e)^{\textsc{cms}}_{\textsc{max}}}\: 
\simeq\: 45.0\: \text{Gev}
\end{cases}
\end{equation}
that in turn yields
\begin{equation}
y_{\textsc{max}} = 
\frac{(E^H_1)_{\cms}^{\textsc{max}}}
{(E^H_1)_{\cms}^{\textsc{max}}+(E^H_2)_{\cms}^{\textsc{min}}} \simeq 0.667
\end{equation}

\noindent Summarizing:
\begin{equation}
\begin{cases}
y_{\textsc{min}} = {\displaystyle 
\frac{(E^H_1)_{\cms}^{\textsc{min}}}
{(E^H_1)_{\cms}^{\textsc{min}}+(E^H_2)_{\cms}^{\textsc{max}}}} \simeq 0.220
\\[4mm]
y_{\textsc{max}} = {\displaystyle 
\frac{(E^H_1)_{\cms}^{\textsc{max}}}
{(E^H_1)_{\cms}^{\textsc{max}}+(E^H_2)_{\cms}^{\textsc{min}}}} \simeq 0.667
\end{cases}
\end{equation}
and it is now possible to give a numerical estimation for the mean 
of the distribution in hadronic energy $a$ depends on:
\begin{equation}
\big< \sqrt{y(1-y)}\: \big>_{\cms}^{\textsc{hera}} = \:
\frac{{\displaystyle \int_{0.220}^{0.667} \sqrt{y(1-y)} \:\ud y}}{0.667-0.220} \:
\: \simeq \: 0.48
~~\text{(cuts considered)}
\end{equation}
or, in our case,
\begin{equation}
\big< M_{\text{X}} \big>_{\cms}^{\textsc{hera}} \: =\: 2E^{\te{had}}_{\te{tot}}\cdot 
\big< \sqrt{y(1-y)}\: \big>_{\cms}^{\textsc{hera}}\: \simeq \: 0.96 \cdot 
E^{\te{had}}_{\te{tot}}
\end{equation}

\section{Final result}
\begin{figure}[!h] 
\begin{center}
\hspace*{-5mm}
\includegraphics[width=0.8\textwidth]{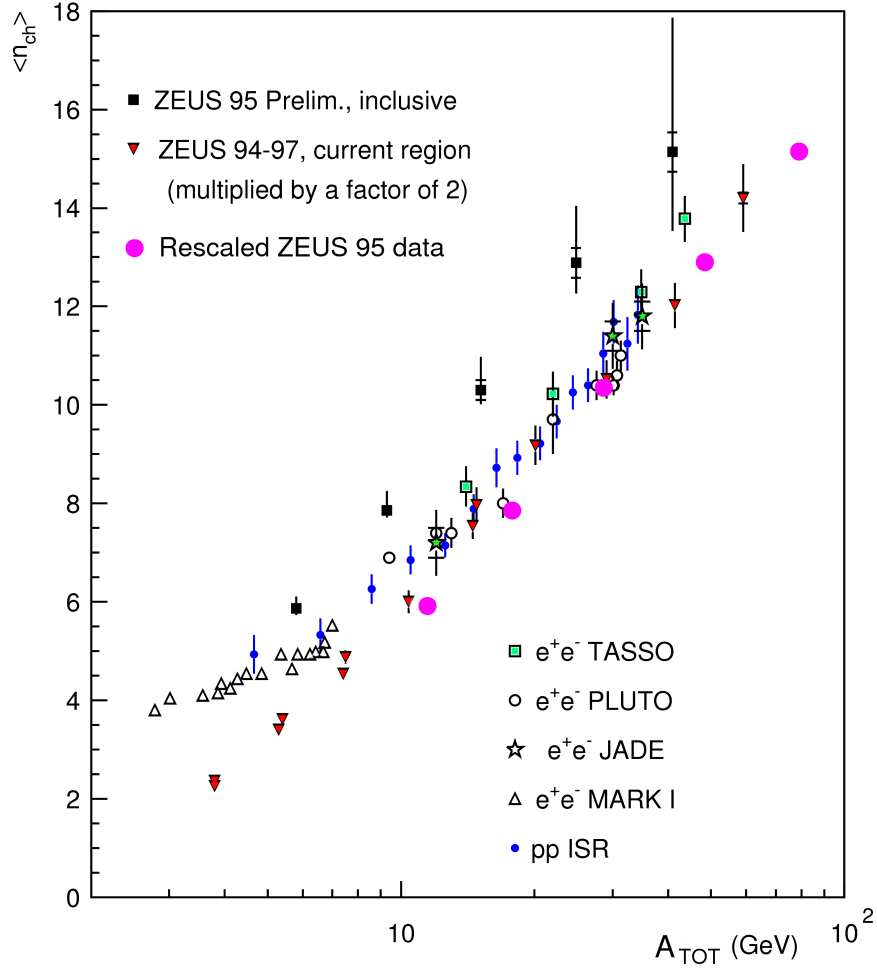}
\caption{The same figure as \ref{osaka} but with DIS data (ZEUS 95) rescaled by the factor 
$a \simeq 2$ and plotted in violet: now the X-axis variable is \atot\!\!. The universality
of \nch is evident, particularly at high energies ($a$ being an asymptotic value).}
\label{rescaled}
\end{center}
\end{figure}

Given the previous results and considerations, we obtain the following
correction factor $a$ to be applied in abscissa in \cite{osaka}:
\begin{equation}
a \: \equiv \: \sqrt{1+\frac{\big< (1-\cos \theta)^2_{\cms} \big>}%
{\big< M_{\textrm{eff}}/(E^H_T)_{\cms} \big>^2}} \: =%
\: \sqrt{1+\frac{2.68}{(0.96)^2}} \: = \: 1.98
\end{equation}
So, the plot will be showing the \nch vs \atot dependence if we multiply DIS data
abscissas by a factor $a$= 1.98. We recall once more that \ee data do not necessitate
any rescaling while for \pp data an estimation of the correction factor is given
below, but we can anticipate that it is negligible. The result is showed in Figure
\ref{rescaled}.

It is evident that, particularly at high energies, DIS data and
other processes data follow the same curve. The quantity \nch
is again a Universality Feature.

\vspace{4mm}

\section{{\boldmath $p\mspace{2mu}$-$p$}  data rescaling}
Recalling that the ISR was balanced and that the full phase space was considered, 
we can conclude that the analyses performed using \eht (the almost totality)
were automatically made in terms of \atot (see equation \ref{eq:aeht}). This means
that those works do not necessitate any correction.

On the contrary, in some other works, the variable \mx was employed
\cite{creation, nc15, nc11, usomx}: these do
need to be corrected for the reasons exposed in the introductory section. This
may seem in contrast with the fact that no differences were found at that time
when data were analysed both with \eht and \mx but it is not so, the reason
being in the low precisions and energies then achievable.

As an example, we estimate the difference between \atot and \mx that
could have been revealed in the early works at ISR where the only condition was
the presence of a leading proton leaving the vertex with 
\begin{equation}
x_{\text{F}} \in [0.4\:,\:0.8]
\end{equation}
from which it follows that
\begin{equation}
E^{\te{had}}_{\te{max}} \:\simeq\: 3\,E^{\te{had}}_{\te{min}}
\end{equation}
and this in turn allows to get the maximum value of $a$ in the ``worst'' situation:
\begin{equation}
\begin{split}
\frac{A_{\text{TOT}}}{M_{\textrm{X}}} &=
\frac{(E^{\te{had}}_{\te{tot}})_{\textsc{cms,lab}}} {M_{\textrm{X}}} \simeq ~~~~~~~
\text{(``worst'' situation)} \\[2mm] & \simeq
\frac{E^{\te{had}}_{\te{max}}+E^{\te{had}}_{\te{min}}}
{2\sqrt{E^{\te{had}}_{\te{max}}E^{\te{had}}_{\te{min}}}}
\simeq  \frac{4\,E^{\te{had}}_{\te{min}}}{2\sqrt{3(E^{\te{had}}_{\te{min}})^2}} =
\frac{2}{\sqrt{3}} \simeq 1.155
\end{split}
\end{equation}
where again the high energies approximation has been used in the
second step.  Thus, in a single event and in the worst
situation, the difference between \atot and \mx could had been
as large as 15\%, but this is a limiting event; when averaging
over all the events the difference becomes $\ll$ 15\% and is
consequently hidden by the large uncertainties that affected
\nch at those times. It is therefore possible to safely employ
ISR \pp data analysed in terms of \mx to make a comparison with
other processes data if a very high precision is not required.


\section{Dependence of {\boldmath $a$ on $\Delta \theta\:$ and
$\Delta y$}}

In this last section the dependence of $a$ on $\Delta \theta\:$
and $\Delta y$ is discussed with the main purpose of showing how
it is improbable for the result obtained in this paper to be
fortuitous.

In fact, changing kinematical conditions (using other values for
$\Delta \theta$ and for $\Delta y$) would bring to different
values of $a$ and the agreement between \ee\!\!, \pp and DIS
achieved in this paper could have not been obtained without
taking a full account of the correct kinematical conditions.

Equation (\ref{eq:ageneral}) shows that $a$ increases with $\theta$
and decreases with the width of the interval in $y$ (centered at
0.5). We examine the situations when $a$ is as high or as low as
possible.  We also check what happens in case of no phase space
reduction.

It is of course safer to vary $\Delta \theta\:$ than $\Delta y$ as from the latter depends
the purity of the DIS sample. Only safe values for the minimum and the maximum of $y$
will be considered.

\noindent
A maximum for $2 \big< \sqrt{y(1-y)}\: \big>$ is obtained by selecting a sample of
balanced events:
\begin{equation}
y = \frac{1}{2} ~\longrightarrow  ~2 \big< \sqrt{y(1-y)}\: \big>_{\textsc{max}} \: = \: 1
\end{equation}
while unbalanced events induce a minimum for $a$. If events with a fast leading proton
and a low energy outgoing positron are chosen (events unbalanced in $E^H$ towards
hemisphere 2)
\begin{equation}
\begin{cases}
x_F \in [0.6\, ;\, 0.8]\\[4mm]
(E'_e)^{\textsc{cms}}_{\textsc{min}} \in [43,7\, ;\, 70] ~\text{Gev}
\end{cases}
\end{equation}
then, in the same way used earlier, we get:
\begin{equation}
\begin{cases}
y_{\te{min}}=0.220 \\[2mm]
y_{\te{max}}=0.431
\end{cases}
\end{equation}
from which
\begin{equation}
\big< \sqrt{y(1-y)}\: \big>_{\textsc{min}} \: = \: 0.464
\end{equation}
a value that may be used to re-evaluate $a$ under modified conditions.
As showed in (\ref{eq:amxpl}) the analytical minimum for $a$ is 1:
\begin{equation}
a \equiv \frac{A_{\text{TOT}}}{M_{\text{eff}}} \le 1
\end{equation}
but it is not possible to reach this value in true experiments as measurements should be
performed in very small and collinear intervals in $\theta\:$ (say $\theta_{\textsc{lab}}
\in [0^0;5^0]$).
Perhaps, a realistic possibility could be $\theta_{\cms}\! \in\! [0^0;90^0]$ that is
$\theta_{\textsc{lab}}\! \in\! [0^0;20^0]$ (a choice that implies a poor
resolution anyway). In such a case 
\begin{equation}
\big< (1-\cos \theta)^2_{\cms} \big> \simeq 0.23
\end{equation}
which yields
\begin{equation}
a_{\textsc{min}} \simeq \sqrt{1+ \frac{0.23}{1}} \simeq 1.1
\end{equation}
that is, in these conditions \atot and \mx would be undistinguishable
in practice, again because of poor resolutions. Such a study would
have never been able to reveal any difference between \atot and \meff
and the universality features would have been evident even using
\meff as it happened in some previous experiments.

In order to get much greater values for $a$ than that obtained in
this work, we should go to the other side of the polar angle range,
imposing, for example,  $\theta_{\cms} \in [150^0;180^0]$. But this
would mean to say  $\theta_{\textsc{lab}} \in [174^0;180^0]$, again a
ridiculous interval. We are thus forced to conclude that
\begin{equation}
a_{\textsc{max}} \gtrsim 2 ~~~(\text{using } 
\big< \sqrt{y(1-y)}\: \big>_{\textsc{min}} \: = \: 0.464)
\end{equation}
and note that this study has already been performed in 
in a kinematical region where $a$ reachs or is very near to its
maximum. 

Finally, it is interesting to see what would have been
obtained by working in the full phase space. Imposing
\begin{equation}
\Delta \theta_{\textsc{lab}} = \Delta \theta_{\cms} = [0^0;180^0] 
\end{equation}
we get
\begin{equation}
\big< (1-\cos \theta)^2_{\cms} \big> = 1.5
\end{equation}
so that
\begin{equation}
a_{\textsc{(full phase space)}} = \sqrt{1+ \frac{1.5}{(0.96)^2}} \simeq 1.62
\end{equation}
a factor implying a difference that is about 40\% smaller than what
has been found in the reduced phase space. This result, combined with
the worse resolution at the two ends of the ZEUS calorimeter, shows
that a full phase space study would have been loosely able to reveal
that smaller difference between \atot and \mx\!.

This discussion shows how the value estimated for $a$ and the
consequent gathering toward the same curve of data points related to
different processes,  has a quite pronounced dependence from the
``contour'' conditions.  As an example, using $a_{\textsc{(full phase
space)}}$ to rescale DIS points in Fig.~\ref{rescaled}, would keep
the mismatch between DIS and other data points evident.

This implies that a possible fortuitousness of the agreement obtained
can be regarded as quite unlikely. It is of course necessary to
perform  direct analyses to check if \atot works well in different
conditions and kinds of experiments, particularly in those involving
the indirect \effl: these processes in particular appear to be the
best candidates to check the geometrical justification for
introducing \atot\!.


\section{Conclusions} 

After reconsidering under the correct geometrical viewpoint the
kinematics of the \effl, a new invariant has been introduced. It
is likely to be the universal Lorentz-invariant quantity
effectively available to multiparticle production and, as such,
the variable that correctly describes the hadronization
processes in any kind of particle collisions. Consequently, it
has been called the Total \aqs\ and denoted with \atot\!.

The universality of \nch has thus been recovered after that
other works \cite{osaka} had claimed its non-universality when
comparing DIS with other kinds of processes and analysing in terms
of the invariant hadronic mass \mx\!\!.

This result definitely rules out \mx as the quantity
representing the so-called ``hadronic energy'' and supports \atot as
its most reasonable successor. In order to exclude other
yet-to-be-proposed possibilities to solve the universality problem,
and eventually adopt \atot to study and describe available and future
experimental results, further analyses concerning other physical
quantities need to be performed.


\section*{Appendix A:\\
A comparison with the generalized parton
distribution approach}

Finally, I would like to make a comparison and discuss similarities
and differences between this work and the generalized parton
distribution (hereafter: GPD) approach to the physics of particles
collisions. For a recent review on the topic see \cite{gpd} and 
references therein. I have found some similarities in the two
viewpoints, but the main difference stays: the incorrect use of
invariant masses to study particles collisions in general and
hadronization in particular.

Of course the GPD formalism digs deep into parton dynamics and enables
to perform explicit calculations concerning exclusive processes and to
study some (otherwise unaccessible) quantities like, for instance,
3-dimensional (transverse) distributions of partons inside the target
hadron (see Impact Parameter Representation, Sect.~3.10 of \cite{gpd})
or helicities. It allows to derive interesting information about the
internal structure of hadrons as well as to calculate form factors for
sub-processes whose experimental study is very difficult or even not
feasible, like graviton-parton interactions or gluonic currents.

The present work is based on 4-dimensional
geometrical-kinematical aspects of the collision processes and
consequently distinguishes longitudinal and transverse
components of involved 4-momenta, just as the GPD formalism
does. This surely is a common point to both approaches but
general results and, in particular, aims are quite different.

In fact the present work concerns with a different aspect of particle
reactions, namely the global and inclusive properties of
multi-hadronic final states, which must show a global universality as
a consequence of QCD universality, as has been extensively discussed
in the paper.

It does not discuss exclusive processes and does not allow to
compute associated amplitudes (even though it tells which is the
correct variable to use to perform such a task), but
concentrates instead on inclusive features of various types of
reactions. We describe the collision as a 2-step process, and
this is another similarity with GPD approach, but there is a
main difference: our 2 steps are ``visible'' in the final
states: the leading particle (or target remnant) is not
considered as a part of the multi-hadronic final state. After
this subtraction, any feature one could think of should
be the same, independently from reacting particles or nominal
energies. I have showed this is indeed the case for the most
essential quantity related to the study and description of
global properties of final states, i.e.~\nch (the mean charged
particle multiplicity).

The GPD approach is unable to recover this universality, the reason
being the use of invariant masses. It forces to account for different
kinematics when comparing processes of different nature (see e.g.
Sect.~6.5.1 of \cite{gpd}), when instead there is no difference at all.
There are no ``diffferent kinematics'', we are simply looking at
things from the wrong point of view (i.e.~using the wrong variables).

The use made of invariant-mass-type quantities, like \mx (which does
not take into account the transverse contribution from the leading
particle) or $W^2$ (which, even worse, does not take into account the
leading particle at all) will never allow to correctly analyse any
kind of experimental data.

Of course GPD could provide very detailed predictions, were the
distinction between the leading particle and the rest of the final
state implemented into the formalism. It would be very interesting to
use the power of GPD to extrapolate results to areas not accessible
to the experiment in order to check and extend the study of
hadronization under the \atot framework toward those events where the
leading particle is not detectable (and thus its subtraction becomes
difficult if not impossible).

At any rate, it is of fundamental importance to abandone at once
-- both for the reason presented here and for experimental
evidences of non-universality (also presented and discussed in
the paper) -- the use of variables like $s$\,, \mx\!, $W^2$ and
similar when their use is evidently improper.

I suspect that many controversial results obtained in the small-$x$
regime (where the leading effect mostly plays its role: high
longitudinal momentum of the leading particle: higher subtraction),
as, for instance, those cited in Section 8 of \cite{gpd}, could be
settled were \atot and related variables used to re-analyse available
or new data.

If the use of \atot (and its derived or derivable quantities) is to
be implemented in the QCD description of both hard partonic
subprocesses and the hadronization phase, I believe that the
longitudinal-transverse separation-oriented GPD formalism, showing
the discussed similarities with the ``\atot approach'', could 
facilitate efforts toward this scope.



\begin{thebibliography}{99}


\bibitem{evid} M.~Basile et al., Phys. Lett. {\bf 92B}, 367 (1980)

\bibitem{nc2} M.~Basile et al., ~N. Cim. {\bf 58A}, N.3, 193 (1980)

\bibitem{nc3} M.~Basile et al., ~Phys. Lett. {\bf 95B}, N.2, 311 (1980)

\bibitem{nc4} M.~Basile et al., ~Phys. Lett. {\bf 99B}, N.3, 247 (1981)

\bibitem{osaka} Zeus Collaboration -- Multiplicity distribution in DIS at HERA. Abstract: 892\\
                Proceedings of the XXXth International Conference on High Energy Physics;
		Osaka, JAPAN \  (08/2000) --- See:
                www-zeus.desy.de/physics/phch/conf/osaka\_paper/QCD/multipdis.ps.gz

\bibitem{creation} The Creation of Quantum Chromo Dynamics and The Effective Energy\\
		   V.~N. Gribov, G.~'t Hooft, G.~Veneziano,
		   V.~F.~Weisskopf -- Edited by L.~N.~Lipatov \\
		   Jointly published by: The university of
		   Bologna and its academy of sciences - The
		   national institute for nuclear physics (INFN)
		   - The italian physics society (SIF), Bologna 1998

\bibitem{nc7} M.~Basile et al., ~Lett. N. Cim. {\bf 30}, N.16, 487 (1981)

\bibitem{nc10} M.~Basile et al., ~Lett. N. Cim. {\bf 32}, N.11, 321 (1981)

\bibitem{hadmes} M.~Basile et al., ~N. Cim. {\bf 66A}, N.2, 129 (1981)

\bibitem{nc9} M.~Basile et al., ~Lett. N. Cim. {\bf 32}, N.7, 210 (1981)

\bibitem{nc8} M.~Basile et al., ~Lett. N. Cim. {\bf 31}, N.8, 273 (1981)

\bibitem{nc5} M.~Basile et al., ~Lett. N. Cim. {\bf 29}, N.15, 491 (1980)

\bibitem{nc15} M.~Basile et al., ~Lett. N. Cim. {\bf 41}, N.9, 293 (1984)

\bibitem{nc11} M.~Basile et al., ~Lett. N. Cim. {\bf 36}, N.10, 303 (1983)

\bibitem{nc14} M.~Basile et al., ~Lett. N. Cim. {\bf 37}, N.8, 289 (1983)

\bibitem{etalead} ``Evidence for $\eta'$ leading production in gluon-induced jets''\\
                  L.~Cifarelli, T.~Massam, D.~Migani, A.~Zichichi\\
		  European Organization for Nuclear Research (1997)  -- CERN papers

\bibitem{usomx} M.~Basile et al., ~N. Cim. {\bf 67A}, N.3, 244 (1982)

\bibitem{zc67} ZEUS Coll., M.~Derrick et al., ~Z. Phys.{\bf C67}, 93-107 (1995)

\bibitem{epjc11} ZEUS Coll., J.~Breitweg et al., ~Eur. Phys. J. {\bf C11}, 251-270 (1999)

\bibitem{pl5} M.~Basile et al., ~Phys. Lett. {\bf 99B}, N.3, 247 (1981)

\bibitem{indpro} M.~Basile et al., ~N. Cim. {\bf 73A}, N.4, 329 (1983)

\bibitem{gpd} M.~Diehl, ~Phys.~Rept.~{\bf388}:41-277, (2003);
hep-ph/0307382

\end{thebibliography}
\end{document}